\documentclass[sn-mathphys-num, 12pt]{sn-jnl}

\usepackage{graphicx}%
\usepackage{multirow}%
\usepackage{amsmath,amssymb,amsfonts}%
\usepackage{amsthm}%
\usepackage{mathrsfs}%
\usepackage[title]{appendix}%
\usepackage{xcolor}%
\usepackage{textcomp}%
\usepackage{manyfoot}%
\usepackage{booktabs}%
\usepackage{algorithm}%
\usepackage{algorithmicx}%
\usepackage{algpseudocode}%
\usepackage{listings}%
\usepackage{array}
\hypersetup{
  colorlinks = true,
  linkcolor =blue,
  anchorcolor = red,
  citecolor = red,
  urlcolor = blue
}
\usepackage{subcaption}
\usepackage{lscape}
\usepackage{float}
\usepackage{booktabs}
\usepackage{comment}
\usepackage{bm}
\usepackage{booktabs}
\theoremstyle{thmstyleone}%
\theoremstyle{thmstyletwo}%
\theoremstyle{thmstylethree}%
\begin{document}

\title[Article Title]{Optimized imaging prefiltering for enhanced image segmentation}

\author*[1]{\fnm{Ronny} \sur{Vallejos}}\email{ronny.vallejos@usm.cl}

\author[1]{\fnm{Felipe} \sur{Osorio}}\email{faosorios.stat@gmail.com}
\equalcont{These authors contributed equally to this work.}

\author[1]{\fnm{Sebastián} \sur{Vidal}}\email{sebastian.vidals@usm.cl}
\equalcont{These authors contributed equally to this work.}

\author[2,3]{\fnm{Grisel} \sur{Britos}}\email{gbritos@unc.edu.ar}
\equalcont{These authors contributed equally to this work.}

\affil*[1]{\orgdiv{Departamento de Matemática}, \orgname{ Universidad T\'ecnica Federico Santa Mar\'ia}, \orgaddress{\street{Avenida España 1680}, \city{Valparaíso}, \country{Chile}}}

\affil[2]{\orgdiv{Facultad de Matemática, Astronomía, Física y Computación}, \orgname{Universidad Nacional de Córdoba}, \orgaddress{\street{Medina Allende s/n, Ciudad Universitaria}, \city{Córdoba},  \country{Argentina}}}

\affil[3]{\orgdiv{Centro de Investigación y Estudios de Matemática}, \orgname{CONICET}, \orgaddress{\street{Medina Allende s/n, Ciudad Universitaria}, \city{Córdoba},  \country{Argentina}}}

%%==================================%%
%% Sample for unstructured abstract %%
%%==================================%%

\abstract{The Box-Cox transformation, introduced in 1964, is a widely used statistical tool for stabilizing variance and improving normality in data analysis. Its application in image processing, particularly for image enhancement, has gained increasing attention in recent years. This paper investigates the use of the Box-Cox transformation as a preprocessing step for image segmentation, with a focus on the estimation of the transformation parameter. We evaluate the effectiveness of the transformation by comparing various segmentation methods, highlighting its advantages for traditional machine learning techniques—especially in situations where no training data is available. The results demonstrate that the transformation enhances feature separability and computational efficiency, making it particularly beneficial for models like discriminant analysis. In contrast, deep learning models did not show consistent improvements, underscoring the differing impacts of the transformation across model types and image characteristics.
}

\keywords{Data transformation, Image enhancement, Image segmentation, Machine learning and deep learning algorithms.}

%%\pacs[JEL Classification]{D8, H51}

%%\pacs[MSC Classification]{35A01, 65L10, 65L12, 65L20, 65L70}

\maketitle

%---------------------------
% Introduction
%---------------------------
\section{Introduction and Motivation}
In 2024, it was 60 years since the Box-Cox transformation was first proposed \citep{Box:1964}. The Box-Cox transformation is a widely used statistical technique for stabilizing variance and making data more closely approximate a normal distribution. Since the seminal paper, power transformations of this type have generated considerable interest, both in theoretical research and in practical applications.

The original form of the Box-Cox transformation takes the form:
\begin{equation}\label{eq:box-cox}
y^{(\lambda)}=\begin{cases} 
\frac{y^{\lambda}-1}{\lambda}, & \text{if} \  \lambda \neq 0,\\
\log(y), & \text{if} \ \lambda =0,
\end{cases}
\end{equation}
and has been designed for positive observations, while  extended versions of the transformation can accommodate negative $y$'s. For instance, Box and Cox proposed the shifted power transformation defined through
\begin{equation}\label{eq:box-cox}
y^{(\lambda)}=\begin{cases} 
\frac{(y+c)^{\lambda}-1}{\lambda}, & \text{if} \  \lambda \neq 0,\\
\log(y+c), & \text{if} \ \lambda =0,
\end{cases}
\end{equation}
where $\lambda$ is the transformation parameter and $c$ is chosen such that $y+c>0$. Other alternatives for handling negative observations can be found in \cite{Manly:1976}, which effectively transforms skewed unimodal distributions into nearly symmetric, normal-like distributions. Subsequently, \cite{John:1980} introduced the so-called modulus transformation. \cite{Bickel:1981} proposed a transformation with unbounded support. Since then, several transformations have been proposed to improve various aspects of the estimated distribution. For example, \cite{Yeo:2000} introduced a new family of power distributions aimed at enhancing normality or symmetry. Practical assessments and evaluations of the transformation can be found in \cite{Os:2010}.

The literature on the estimation of $\lambda$ is extensive. Although Box and Cox provided an estimate for $\lambda$ that maximizes the profile likelihood of the transformed data, their estimation was restricted to a grid of $\lambda$ values lying within a confidence interval obtained via a likelihood ratio test. This proposal has been a subject of debate in the literature \cite{Riani:2023}, with some authors opting not to follow this recommendation \cite{Chen:2002}. The impact of plugging in an estimate of $\lambda$ on the variance of linear model parameters has been examined in \cite{Mc:2002} and \cite{Proietti:2009}.

The application of the Box-Cox transformation in image processing is relatively recent (\cite{Cheddad:2020}). The general concept of transforming pixel intensity values is well established, particularly for adjusting contrast and brightness in image enhancement. In a general framework, if $f(u,v)$ represents the intensity $f$ measured at  location $(u,v)$, then a transformed image, denoted as $g(u,v)$, is defined as
\begin{equation}\label{Eq:Trans}
g(u,v)=T\left[ f(u,v)\right],  
\end{equation}
where $T$ is a transformation operator to $f$, within a neighborhood of $(u,v)$ \cite{Gonzalez:2011}. Several transformations used in the literature are special cases of Equation  \eqref{Eq:Trans} (see, for instance, \cite{Maurya:2022}). Recently, the Box-Cox transformation has been employed as a preprocessing tool for two-dimensional data (digital images), with a focus on analyzing its effects \cite{Cheddad:2020}. In particular, the study aimed to evaluate improvements in visual quality and image enhancement resulting from its application.

In this work, we also explore the Box-Cox transformation within the framework of image enhancement; however, we distinguish our approach from that of \cite{Cheddad:2020} in two key aspects. First, we revisit the estimation of $\lambda$
to analyze the classification rate as a function of the estimated $\lambda$.  Second, our focus on image enhancement is specifically aimed at improving image segmentation. To achieve this, we compare several segmentation methods after applying the Box-Cox transformation as a preprocessing filter. 
Numerical experiments with real images demonstrate that the preprocessing filter is particularly effective for the discriminant analysis technique, especially when there is no data available to train other competitive approaches. It also underscores the importance of accurately estimating $\lambda$.

The paper is structured as follows: Section \ref{sec:Methods} describe the methodology used in this study: the use of prefiltering transformations as a previous step in the modeling process, some precision and concordance measures to assess the evaluation of the technique, and the expected improvement of the prefiltering on image segmentation. Section \ref{sec:Results} presents the results obtained for neural network and other machine learning models, and describes the estimation of $\lambda$ and its impact on image segmentation. The paper finishes with a discussion and an outline of future research problems. The description of the methods used in this paper is relegated to Appendix \ref{secA1}, extra images for the results section are in  Appendix \ref{secA2}. An additional segmentation example is provided in Web Appendix A of the supplementary material associated with this paper.

%---------------------------
% Methods
%---------------------------
\section{Methods}\label{sec:Methods}

%---------------------------
%Prefiltering
%---------------------------
\subsection{Image Prefiltering and   $\lambda$ Estimation}\label{subsec:PFI}
In image processing, contrast enhancement is a fundamental step aimed at improving visualization and image quality, particularly in cases where important details may be hidden due to a limited dynamic range of grayscale levels. The following introduces several existing prefiltering methods for image enhancement.

Histogram stretching is an image processing technique that enhances image contrast by expanding the range of pixel intensities through histogram normalization. Suppose an image has an intensity level range of 
$[f_{\min},f_{\max}]$, where $f_{\min}$ and  $f_{\max}$ 
represent the minimum and maximum intensity levels, respectively. The goal of histogram stretching is to map this range to a new intensity level range, typically 
$[g_{\min},g_{\max}]$, covering the full possible intensity range of the image. Using the notation of \eqref{Eq:Trans}, the linear function  that performs this histogram stretching can be expressed as
\begin{equation}\label{hist}
g(u,v)=\frac{(f(u,v)-f_{\min})}{(f_{\max}-f_{\min})} \times (g_{\max}-g_{\min})+g_{\min},
\end{equation}
where $g_{\min}$ and $g_{\max}$ represent the minimum and maximum intensity levels of the transformed image. 
The histogram stretching process expands the range of intensities to cover the full spectrum, thereby enhancing the image contrast. A recent review on pixel-based enhancement techniques can be found in \cite{Chris:2025}.

Gamma correction is a nonlinear transformation commonly used to adjust the brightness of an image. It modifies pixel values to better align with the way the human eye perceives light. The transformation is defined as follows 
\begin{equation}\label{Eq:gaamma}
g(u,v)=c f(u,v)^{\gamma},
\end{equation}
where $c$ and $\gamma$ are positive constants.
When $\gamma<1$ pixel intensities in dark regions of the image are increased, while bright pixels remain largely unchanged. Conversely, when $\gamma>1$, pixel intensities in bright regions are reduced, while dark pixels remain mostly unaffected. This adjustment is particularly useful for images that are overly bright and require greater contrast in lighter areas. If 
$\gamma=1$, no correction is applied.

Let $\mathcal{F}$ be a color image in the red-green-blue (RGB) space, defined as
\begin{equation}\label{Eq:space}
\mathcal{F}=\{f(u,v): f(u,v)=(R(u,v), G(u,v), B(u,v)), u=1,\ldots,U, v=1,\ldots,V \},   
\end{equation}

where $U$ and $V$ represent  the image  dimensions. An image in $\mathcal{F}$ can also be converted to grayscale using the transformation
\begin{equation*}
f^{\prime}(u,v)=0.299\cdot R(u,v)+0.587 \cdot G(u,v)+ 0.114 \cdot B(u,v).
\end{equation*}

This grayscale representation has been shown to be effective in separating high-frequency signals from chromatic components, which are blended in the RGB space \cite{Cheddad:2020}. In this context, let $f^{\prime \prime}(u,v)^{(\lambda)}$ denote the Box-Cox transformation of $f^{\prime}(u,v)$. The corresponding histogram stretching transformation of $f^{\prime \prime}(u,v)^{(\lambda)}$, denoted $g^{\prime \prime}(u,v)^{(\lambda)}$,  can then be defined as:

\begin{equation}\label{Eq:finalT}
g^{\prime \prime}(u,v)^{(\lambda)}=\frac{(f^{\prime \prime}(u,v)^{(\lambda)}-f^{\prime \prime}_{\min})}{(f^{\prime \prime}_{\max}-f^{\prime \prime}_{\min})} \times (g^{\prime \prime}_{\max}-g^{\prime \prime}_{\min})+g^{\prime \prime}_{\min},
\end{equation}
where $f^{\prime \prime}_{\min}$ and $f^{\prime \prime}_{\max}$ represent the minimum and maximum values of $f^{\prime \prime}(u,v)^{(\lambda)}$, and the interval $[g^{\prime \prime}_{\min}, g^{\prime \prime}_{\max}]$ defines the desired range for the image $g^{\prime \prime}(u,v)^{(\lambda)}$.

It is worth noting that this transformation combines two previously described image enhancement techniques: histogram stretching and a particular case of gamma correction (Box-Cox transformation). In this case, the parameter $\lambda$ is optimally selected by maximizing the likelihood function of the transformed data. To perform this estimation, it is convenient to represent the image in vector form. Given that a grayscale image contains $n = U \times V$ pixels, we define its vectorized version as $\bm{y} \in \mathbb{R}_n^+$, where each entry $y_i$ corresponds to a nonnegative pixel intensity. This  representation allows us to apply standard statistical modeling techniques for random vectors.

 Let $\bm{y}^{(\lambda)} = (y_1^{(\lambda)},\ldots, y_n^{(\lambda)})^{\top}$ denote the Box-Cox-transformed version of $\bm{y}$. We assume a linear Gaussian model for the transformed data:
\[
\bm{y}^{(\lambda)} \sim \mathcal{N}(\mathbf{A}\boldsymbol{\theta}, \sigma^2 \mathbf{I}),
\]
where $\mathbf{A}$ is a full-rank design matrix of dimension $n \times p$.
By the transformation theorem, the log-likelihood of $\bm y$ is given by
$$\ell(\lambda, \bm y)=\frac{n}{2}\log(\sigma^2)-\frac{1}{2 \sigma^2}||\bm y^{(\lambda)}-\bm A \bm \theta||^2-(\lambda-1)\sum_{i=1}^n \log(y_i),$$
from which the estimators $\bm{\theta}$ and $\sigma^2$ are obtained as
$$\widehat{\bm \theta}=(\bm A^{\top} \bm A)^{-1} \bm A^{\top} \bm y^{(\lambda)},\  \text{and} \ \widehat{\sigma}^2=\frac{1}{n}||\bm y^{(\lambda)}-\bm A \widehat{\bm \theta}||. $$
Finally, the estimation of $\lambda$ is obtained from the profile likelihood function
$$\ell(\lambda)=\frac{n}{2}\log \left( \frac{1}{n} ||\bm y^{(\lambda)}-\bm A \bm \theta||^2\right)+(\lambda-1)\sum_{i=1}^n \log(y_i).$$

We emphasized that in this paper, we do not address segmentation techniques for color images; instead, we focus exclusively on segmentation based on grayscale images. Nonetheless, many methods initially developed for grayscale segmentation have been successfully adapted for use with color images \cite{Du:2007}. Consequently, the development of color image segmentation algorithms must consider the specific properties and challenges associated with color data \cite{Garcia:2018}.
%%%%%%%%%%%%

%---------------------------
% Performance and Concordance
%---------------------------
\subsection{Performance and concordance measures}
\label{sec:Performance}

In this section, we provide a concise review of several indices derived from the confusion matrix that have been employed to assess the quality of segmentation and classification techniques. These indices will be utilized in Section \ref{sec:Results}.

Assume that from a confusion matrix we have obtained the indices: True Positive (TP), False Positive (FP),  False Negative (FN), and True Negative (TN). Then
\begin{align*}
\text{Precision}&=\frac{TP}{TP+FP},\\
\text{Recall}&=\frac{TP}{TP+FN},\\
\text{Accuracy}&=\frac{TP+TN}{TP+TN+FP+FN},\\
F_1 \text{score}&=\frac{\text{Precision}\cdot \text{Recall}}{\text{Precision}+ \text{Recall}}.
\end{align*}
In the context of image classification and segmentation, overlap metrics are used to assess how well a segmented image matches the ground truth mask or true segmentation. These metrics quantify the intersection and similarity between the pixels of the predicted regions and the actual regions of interest in the image.
Specifically, we introduce the IoU and Dice indices,  which are defined based on two sets: $A$, representing the predicted pixels, and 
$B$, representing the true pixels. These indices are then formulated as follows \citep{Jak:2021}:
\begin{align*}
\text{IoU}&=\frac{|A \cap B|}{|A \cup B|},\\
\text{Dice}&=\frac{2 |A \cap B|}{|A| + |B|},
\end{align*}
where $|\cdot|$ denotes the cardinality of a set. The IoU and Dice indices, the closer they are to 1, the greater the overlap between the two regions, indicating a more accurate segmentation.

The Kappa coefficient defined below is introduced as a corrected and normalized measure of agreement \cite{Sim:2005}, used to measure the degree of agreement between two algorithms or classifiers, taking into account the possibility that some of the coincidence occurs by chance.

The index is defined through
\begin{equation}
    \kappa=\frac{p_0-p_e}{1-p_e},
\end{equation}
where $p_0$  represents the proportion of observed agreement between the classifier and the reference, and $p_e$ denotes the expected agreement by chance. As with standard agreement measures, $\kappa \in [-1,1].$ Values of $\kappa$ close to 1 or -1 indicate a substantial level of agreement \cite{Landis:1977}.

%---------------------------
%Image Segmentation
%---------------------------
\subsection{Improving image segmentation algorithms}
Image segmentation, a prominent research focus in image processing and computer vision, involves partitioning an image into meaningful, non-overlapping regions and plays a crucial role in natural scene understanding \cite{Pal:1993}. Despite significant progress over the years, challenges remain in feature extraction and model design. 

These difficulties arise in part because image segmentation is interpreted differently across various application domains, and also due to the diversity of textures, which complicates the development of a single segmentation algorithm suitable for all image types. A recent review on image segmentation, along with a classification of the various techniques used to address this problem, can be found in \cite{Yu:2023}.

The proposed improvement is grounded in the premise that effective prefiltering  enhances image segmentation, particularly in scenarios with limited data for methods that rely on extensive training. In such cases, traditional approaches that assume specific data distributions and stable variance offer a viable alternative that warrants renewed attention.

\begin{figure}[htp]
\centering
\includegraphics[scale=0.54]{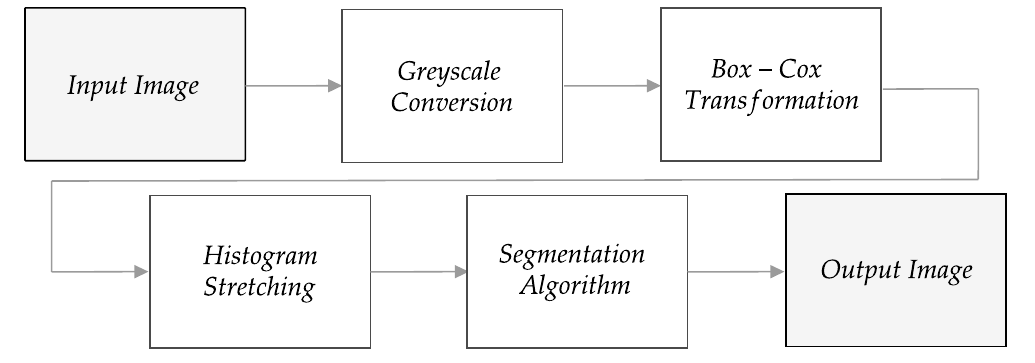}
\caption{Block diagram representation of the proposed methodology for improving image segmentation.}
\label{Fig:Scheme}
\end{figure}

The method works for a color image, which is transformed to a grey scale.  The Box-Cox transformation is applied then using an optimal value of $\lambda$. Histogram stretching is applied to the resulting image to ensure that the whole range of grey intensities is covered. Finally, an image segmentation algorithm is performed over the prefiltered image as is illustrated in Figure \ref{Fig:Scheme}.

%\begin{landscape}
%\begin{center}
%\end{center}
%\end{landscape}
It should be noted that, after the segmented image is obtained, an evaluation of segmentation quality using appropriate metrics is necessary. Alternatives for selecting a suitable quality metric are discussed in Section \ref{sec:Performance}.

In Section \ref{sec:Results}, various aspects of the proposed methodology are examined through numerical experiments. Specifically, we analyze the impact of $\lambda$ estimation on the effectiveness of the methodology and compare the performance of statistical segmentation methods with that of machine learning-based segmentation algorithms. We categorize the segmentation methods employed in this study into two groups. The first group consists of methods that require training data to generate accurate predictions, such as a Semantic Segmentation Model (DeepLab), Fully Convolutional Networks (FCN), and U-Net. The second group includes methods commonly used in machine learning, such as Support Vector Machine (SVM), Light Gradient-Boosting Machine (LightGBM), K-Nearest Neighbor (K-NN), Linear Discriminant Analysis (LDA), and Quadratic Discriminant Analysis (QDA). Further details on the classification rules associated with these methods can be found in Appendix \ref{secA1}.
%---------------------------
% Results
%---------------------------
\section{Results}\label{sec:Results}
This section investigates the effect of applying the Box-Cox transformation, with estimated parameter $\lambda$, as a preprocessing step for image segmentation tasks. We conduct experiments on multiple real-world datasets—including satellite imagery, crack detection, and lunar surface segmentation—to evaluate its impact on both classical and deep learning-based segmentation models. Using standard performance metrics, we analyze how modifying the distribution of gray levels in input images influences segmentation accuracy. The objective is to determine whether the Box-Cox transformation can consistently enhance segmentation outcomes by improving the input data representation prior to model training.

\subsection{Neural Network Models}\label{sec:NNM}
In order to explore the performance of the neural network approach to image segmentation using the Box-Cox transformation as a prefiltering procedure, we consider two images from the {\it kaggle} database \footnote{https://www.kaggle.com}. This dataset provides a collection of 2,841 satellite images of 512 × 512 pixels captured by the Sentinel-2 satellite, focusing on bodies of water such as lakes, rivers, oceans, and other types of water masses. The images are labeled and organized to facilitate the training of machine learning models, particularly for tasks related to the classification, segmentation, and detection of bodies of water from space. 

For this study, the satellite image dataset of water surfaces was used to train three neural network models: DeepLabV3, U-Net, and FCN. ResNet50 was used as the encoder for DeepLabV3 and U-Net, while a custom encoder, based on a series of convolutional and pooling layers, was designed for FCN. The images were split into 2,023 for training, 11 for validation, and 664 for testing. The selection of the best hyperparameters was carried out through cross-validation, simultaneously with model training.

To increase the quantity and diversity of the training data without collecting new images, data augmentation techniques such as flipping, rotation, cropping, and scaling were applied to the available images. After training all models and tuning the hyperparameters, the test set images were segmented both before and after the Box-Cox transformation. Finally, given that neural networks produce probabilistic rather than deterministic outputs, a threshold of 0.5 was used to classify pixels as either water surfaces or non-water areas.

%se utilizó la media aritmética
Table \ref{table:resred} shows that the U-Net method without the Box-Cox transformation achieves the best results in terms of Dice (68. 81\%) and IoU (56. 16\%), suggesting a better overall segmentation. However, when the Box-Cox transformation is applied, U-Net experiences a decline in all metrics. Similarly, the FCN method with Box-Cox also shows a drop in all indicators compared to the untransformed images. Finally, DeepLab, with and without Box-Cox, exhibits the lowest performance, particularly in IoU and Dice, indicating that the Box-Cox transformation does not seem to improve its performance. Although this does not highlight the improvement produced by the prefiltering technique, this outcome is expected, as neural networks process a vast amount of information during training, significantly influencing their behavior. As a result, prefiltering does not make a noticeable difference.

 \begin{table}[h]
\centering
\scalebox{1.3}{
\begin{tabular}{lccc}
\toprule
Method              & IoU     & Dice    & Accuracy \\ \midrule
DeepLab             & 44,77 & 34,45 & 76,72  \\ 
 DeepLab Box-Cox & 20,90 & 28,91 & 69,34  \\ 
FCN                 & 49,91 & 65,27 & 70,36  \\ 
FCN Box-Cox     & 49,64 & 63,20 & 74,91  \\ 
U-net               & 56,16 & 68,81 & 74,48  \\ 
U-net Box-Cox   & 47,76 & 61,13 &  68,27  \\ \bottomrule
\end{tabular}
}
\caption{Quality assessment (percentage of correct segmentation) of deep learning models.}
\label{table:resred}
\end{table}

\begin{figure}[H]
\centering
\begin{subfigure}[b]{0.19\linewidth}
\includegraphics[width=\linewidth, height=4cm]{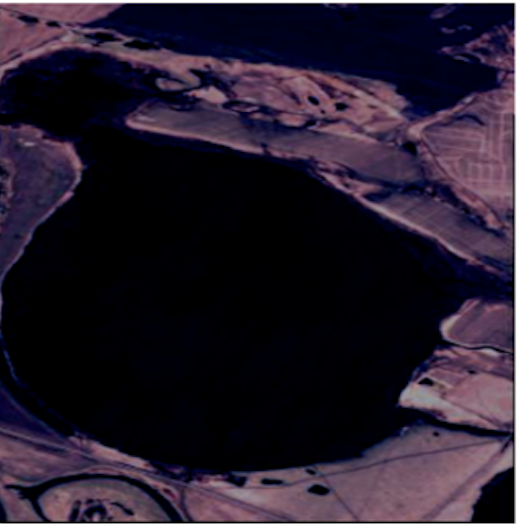}
\caption{}
\label{fig:or}
\end{subfigure}
\begin{subfigure}[b]{0.19\linewidth}
\includegraphics[width=\linewidth, height=4cm]{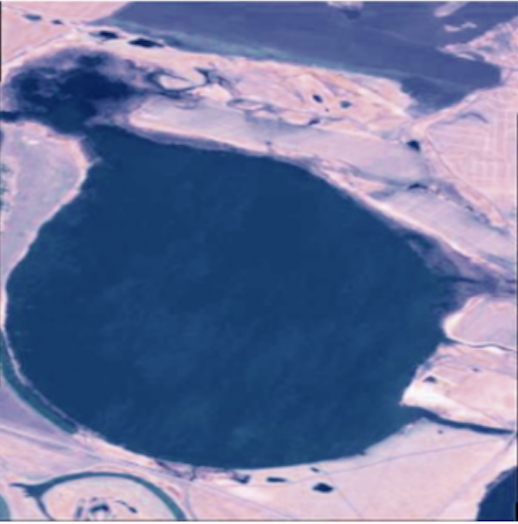}
\caption{}
\label{fig:mask}
\end{subfigure}
\begin{subfigure}[b]{0.19\linewidth}
\includegraphics[width=\linewidth, height=4cm]{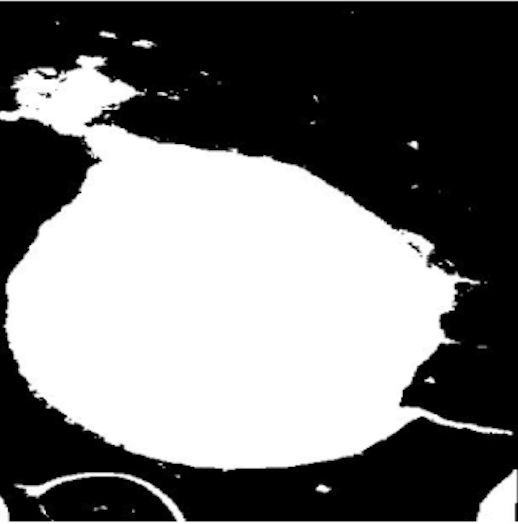}
\caption{}
\label{fig:mask}
\end{subfigure}
\begin{subfigure}[b]{0.19\linewidth}
\includegraphics[width=\linewidth, height=4cm]{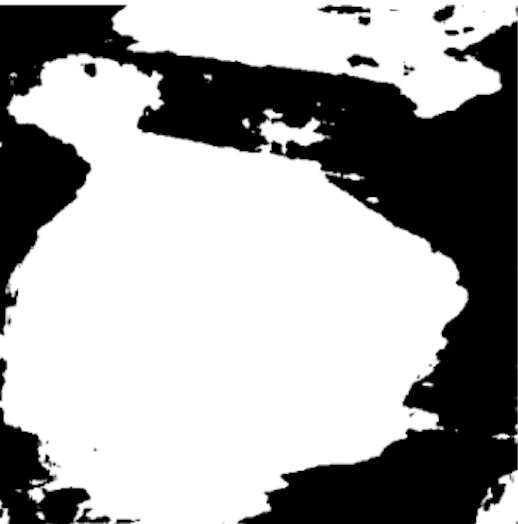}
\caption{}
\label{fig:pred_or}
\end{subfigure}
\begin{subfigure}[b]{0.19\linewidth}
\includegraphics[width=\linewidth, height=4cm]{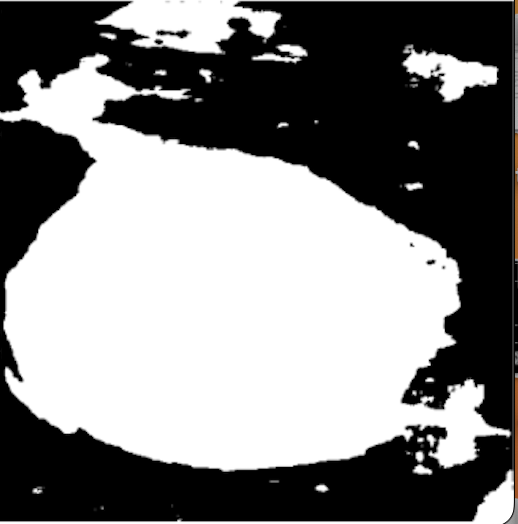}
\caption{}
\label{fig:pred_trans}
\end{subfigure}
\caption{Predictions using the U-net method. (a) Original image from the kaggle database;  (b) Box-Cox transformation of (a); (c) Grey intensity level mask of (a); (d) Prediction using the original image as  input;  (e) Prediction using the Box-Cox transformation image as  input.}
\label{fig:pred_red}
\end{figure}

\begin{figure}[H]
\centering
\begin{subfigure}[b]{0.19\linewidth}
\includegraphics[width=\linewidth, height=4cm]{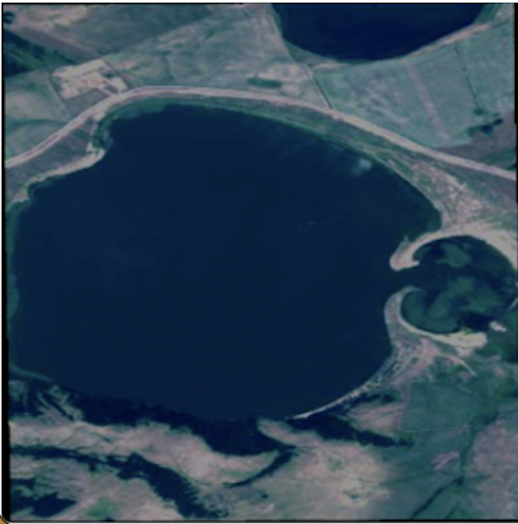}
\caption{}
\label{fig:or}
\end{subfigure}
\begin{subfigure}[b]{0.19\linewidth}
\includegraphics[width=\linewidth, height=4cm]{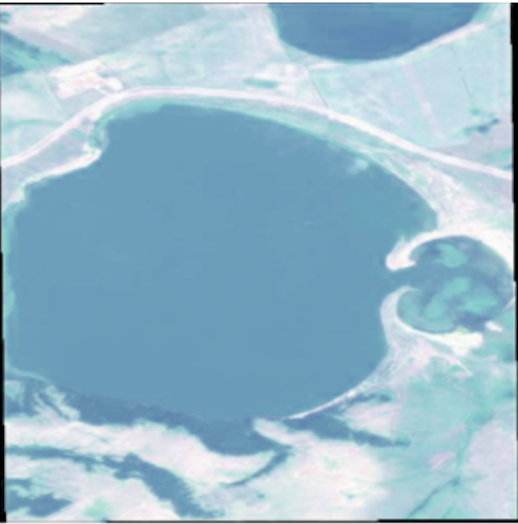}
\caption{}
\label{fig:mask}
\end{subfigure}
\begin{subfigure}[b]{0.19\linewidth}
\includegraphics[width=\linewidth, height=4cm]{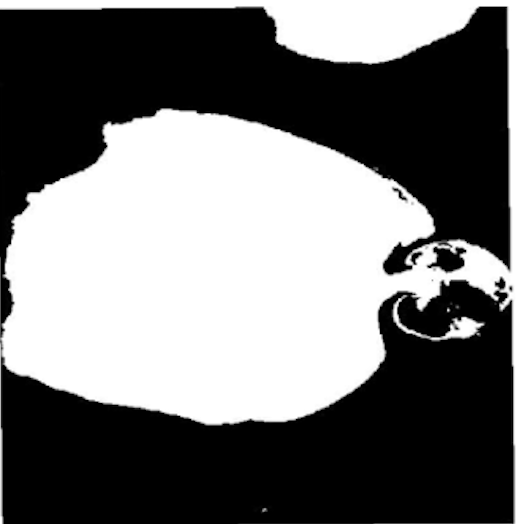}
\caption{}
\label{fig:mask}
\end{subfigure}
\begin{subfigure}[b]{0.19\linewidth}
\includegraphics[width=\linewidth, height=4cm]{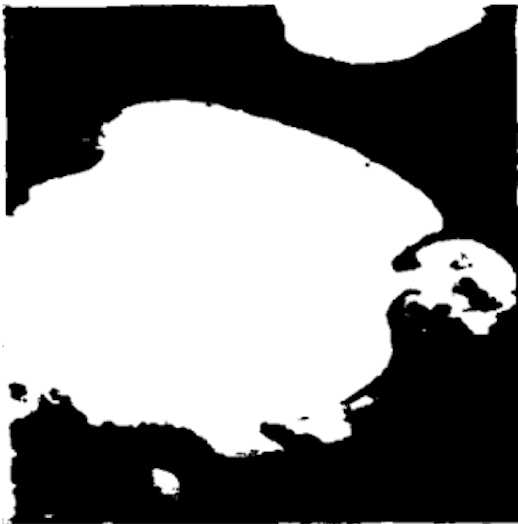}
\caption{}
\label{fig:pred_or}
\end{subfigure}
\begin{subfigure}[b]{0.19\linewidth}
\includegraphics[width=\linewidth, height=4cm]{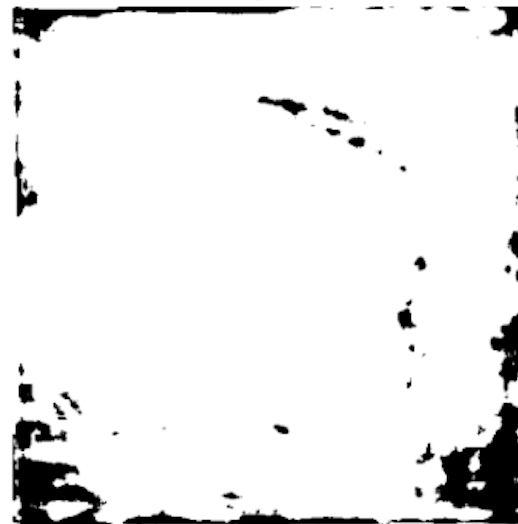}
\caption{}
\label{fig:pred_trans}
\end{subfigure}
\caption{Predictions using the U-net method. (a) Original image from the kaggle database;  (b) Box-Cox transformation of (a); (c) Grey intensity level mask of (a); (d) Prediction using the original image as  input;  (e) Prediction using the Box-Cox transformation image as  input.}
\label{fig:pred_red2}
\end{figure}

To visually assess the performance of the deep learning methods, two sets of images are presented in Figures \ref{fig:pred_red} and \ref{fig:pred_red2}. In Figure \ref{fig:pred_red}, it is evident that the Box-Cox transformation enhances both the visual quality of the image and the segmentation accuracy. However, in Figure \ref{fig:pred_red2}, the transformation has an adverse effect, resulting in poorer image segmentation compared to the original prediction.

\subsection{Machine Learning Models}\label{MLM}
In this section, we present two numerical examples to illustrate that, for the machine learning models used, prefiltering plays a key role in the final segmentation of an image. 

\begin{figure}[h]
    \centering
    \begin{subfigure}[b]{0.3\linewidth}
    \includegraphics[width=\linewidth]{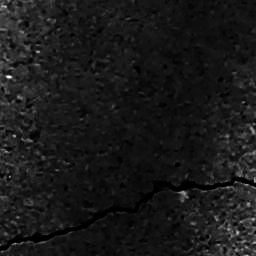}
    \caption{}
    \end{subfigure}
    \begin{subfigure}[b]{0.55\linewidth}
    \includegraphics[width=\linewidth]{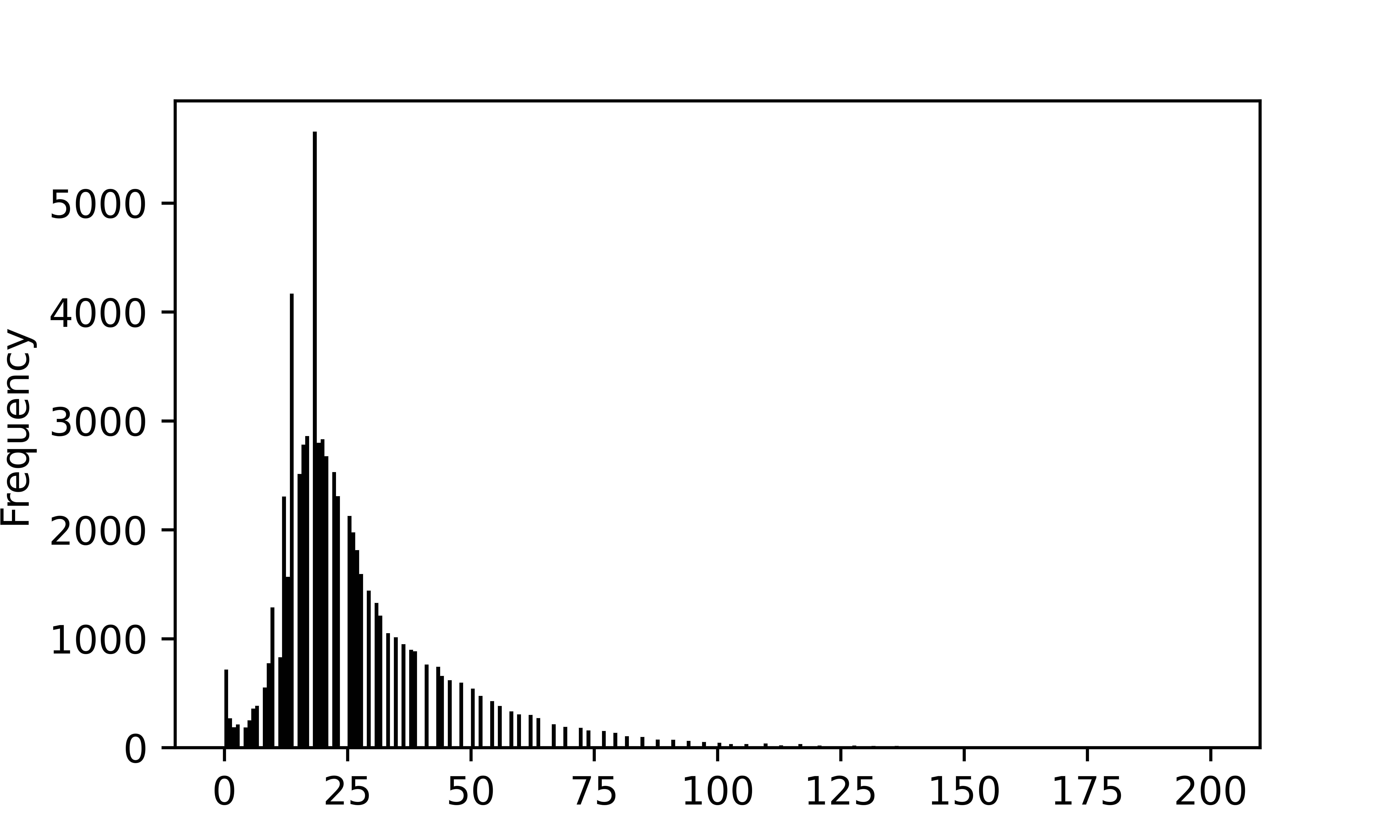}
    \caption{}
    \end{subfigure}\\
    \begin{subfigure}[b]{0.3\linewidth}
    \includegraphics[width=\linewidth]{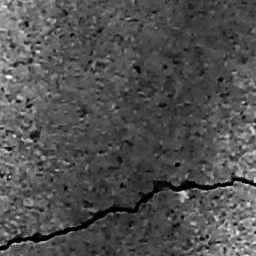}
    \caption{}
    \end{subfigure}
    \begin{subfigure}[b]{0.55\linewidth}
    \includegraphics[width=\linewidth]{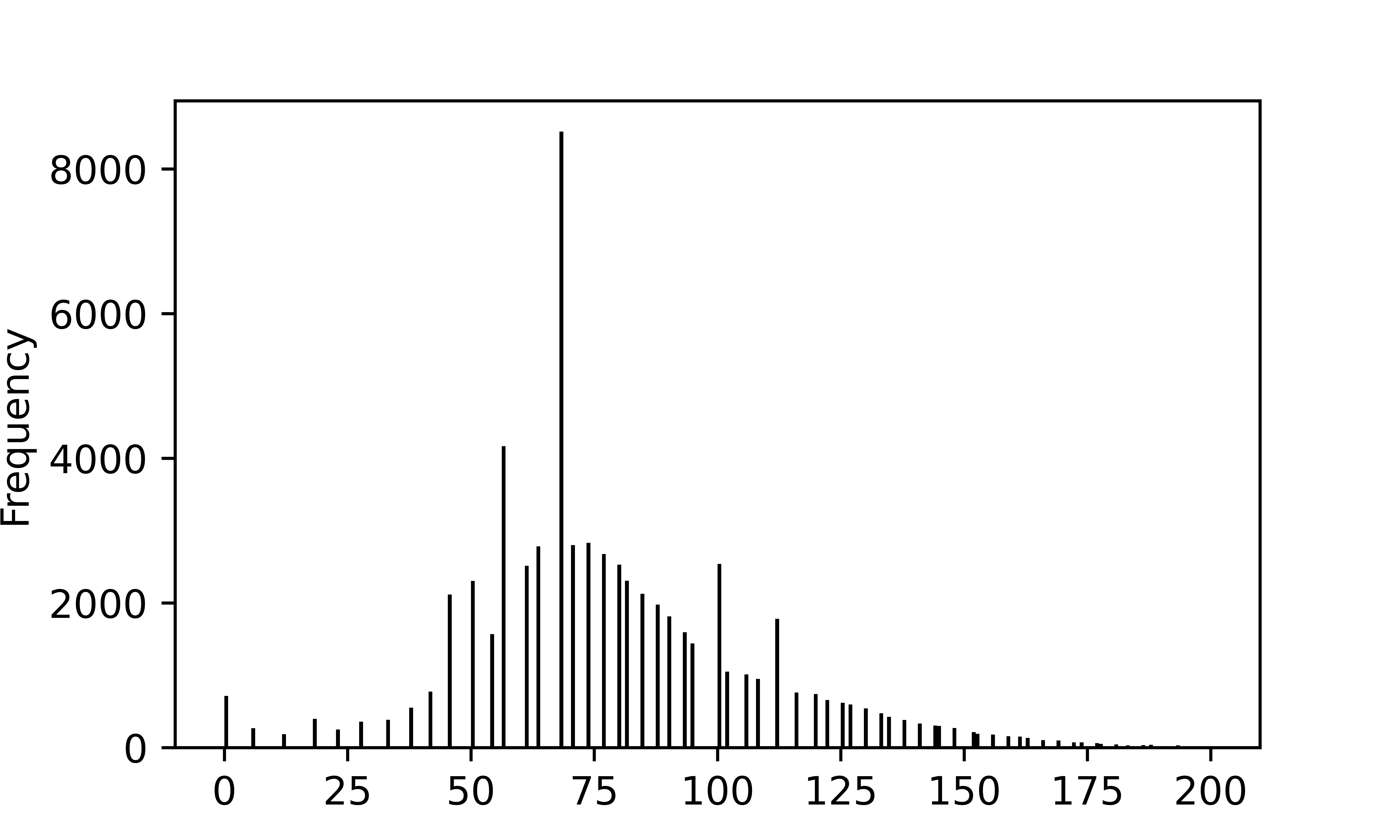}
    \caption{}
    \end{subfigure}
    \caption{Crack image. (a) Original image; (b) Histogram of the original image; (c) Image after applying the Box-Cox transformation; (d) Histogram of the transformed image.}
    \label{fig:cracking}
\end{figure}

A low-contrast 
$227\times 227$
 image of a concrete surface was selected \cite{Oz:2018} to apply the Box-Cox transformation and compare the histograms before and after the transformation. Figure \ref{fig:cracking} illustrates the effect of the transformation on an image of a crack. In the first image (a), the pixels are predominantly concentrated in low-intensity values, resulting in an overall low contrast. This is confirmed by the histogram (b), which exhibits a pronounced left-skewed asymmetry, indicating a significant accumulation of low-intensity values. After applying the transformation, as shown in image (c), there is a noticeable improvement in contrast, clearly highlighting the most relevant feature—the crack. This enhancement is reflected in the post-transformation histogram (d), which displays a more uniform and symmetric distribution of intensities, confirming an increased dynamic range and improved visual representation of the image features.

The methods SVM, LightGBM, LDA, KNN and QDA were applied to the crack image shown in Figure  \ref{fig:cracking}(a) to produce a segmentation before and after the Box-Cox transformation. In general terms, applying the transformation improved the segmentation performance of the LDA and QDA methods by approximately 3\% (Table \ref{table:resgrieta}). While the increase in LDA performance is not particularly significant in terms of precision, recall, and F1 score, the normalized confusion matrix before and after the transformation, shown in Tables \ref{table:antes_despues}, indicates that the transformation enhances the model's ability to identify cracks more effectively.

  \begin{table}[h]
\centering
\scalebox{1.3}{
\begin{tabular}{lccc}
\toprule
Method                    & Precision     & Recall        & $F_1$ score    \\ \midrule
SVM                       & 99          & 99          & 99         \\ 
SVM Box-Cox      & 99 & 99 & 99 \\ 
LightGBM                  & 99          & 99         & 99          \\ 
LightGBM Box-Cox & 99 & 99& 99 \\ 
LDA                       & 96          & 98          & 97         \\ 
LDA Box-Cox      & 99 & 99 & 99 \\ 
K-NN                      & 99          & 99         & 99         \\ 
K-NN Box-Cox     & 99 & 99 & 99 \\ 
QDA                       & 96          & 98          & 97       \\
QDA Box-Cox      & 99 & 99 & 99 \\ \bottomrule
\end{tabular}
}
\caption{Quality assessment (percentage of correct segmentation) of machine learning models.}
\label{table:resgrieta}
\end{table}

\begin{table}[h]
\centering
\scalebox{1.3}{
\begin{tabular}{ccccc}
\toprule
 & \multicolumn{2}{c}{Before Transformation} & \multicolumn{2}{c}{After Transformation} \\
\cmidrule(lr){2-3} \cmidrule(lr){4-5}
        & Concrete & Crack & Concrete & Crack \\
\midrule
Concrete & 100 & 0  & 99  & 1    \\ 
Crack    & 100 & 0  & 34  & 66   \\
\bottomrule
\end{tabular}
}
\caption{Percentages of normalized confusion matrices for the segmentation results obtained using LDA before and after applying the Box-Cox transformation.}
\label{table:antes_despues}
\end{table}

The segmentation results align with the findings reported in this study. The processed crack image reveals that the original segmentation yielded entirely black images. However, after the transformation was applied, a significant improvement was observed for the LDA and QDA methods. This enhancement underscores the effectiveness of the prefiltering technique, which leverages the distributional assumptions of these methods.

Although SVM and K-NN exhibited comparable performance in segmenting the crack image both before and after the transformation, an important consideration is the training time required for each method. Table \ref{table:times} presents the time needed to estimate the hyperparameters for each approach. In particular, techniques such as LDA and QDA required substantially less training time than the others.

\begin{table}[h]
\centering
\scalebox{1.3}{
\begin{tabular}{cccccc}
\toprule
Method & SVM & LightGBM & LDA & K-NN & QDA\\ \midrule
Time (s) & 109.8 & 117 & 15 & 445.2 & 0.72 \\ \bottomrule
\end{tabular}
}
\caption{Training times for the estimation of hyperparameters for each of the methods used in the segmentation of the crack image.}
\label{table:times}
\end{table}

Here, we present a second example to illustrate the diversity of images analyzed in this study.

Conducting any type of machine learning experiment on lunar images is generally challenging due to their scarcity and lack of annotations. The goal of this dataset, created by the Space Robotics Group at Keio University \cite{Petrakis:2024}, is to provide the public with a collection of realistic yet artificial lunar landscapes that can be used to train rock detection algorithms. These trained algorithms can then be tested on real images of the Moon or other rocky terrains.

In Figure \ref{fig:luna_modelos}, the mask contains five labels: small rocks (green), large rocks (blue), lunar surface (black), and sky (red). However, for the purposes of this study, the segmentation will be simplified to consider only three classes: sky, large rocks, and lunar surface. Table \ref{table:metodos_luna} presents the performance metrics of the different models. In general, no significant changes in model performance are observed after applying the Box-Cox transformation.

When evaluating the confusion matrix before and after the transformation for the QDA model (Table \ref{table:antes_despues_roca}), it is observed that prior to the transformation, the model correctly classified 91\% of the pixels corresponding to the sky, while the remaining 9\% were incorrectly classified as surface. After applying the transformation, there was a slight decrease in accuracy for the sky class, dropping to 89\%. However, for the rock class, the transformation significantly improved the model's performance, going from a completely incorrect classification to a correct classification of 79\% of the pixels corresponding to rock.

Table \ref{table:times_rock} presents the training and validation times for lunar rock image segmentation, a more complex scenario than the previous one due to the presence of three categories: surface, rock, and sky. The increased complexity of the problem is reflected in a considerable rise in training times for the SVM methods, whereas for LDA and QDA, the times remained practically constant. These results highlight the importance of considering both accuracy and time efficiency when selecting a method for image segmentation in more complex contexts.

All scripts, trained weights, and images used in this paper are publicly available at  
\texttt{https://github.com/sebastianvidal92/BCI\_segmentation}.

\subsection{Estimation of $\lambda$}\label{sec:Lambda}

The estimates of $\lambda$ used in the results presented in Sections \ref{sec:NNM} and \ref{MLM} were obtained using the methodology described in \ref{subsec:PFI}.

\begin{figure}[h!]
    \centering
    \includegraphics[scale=0.55]{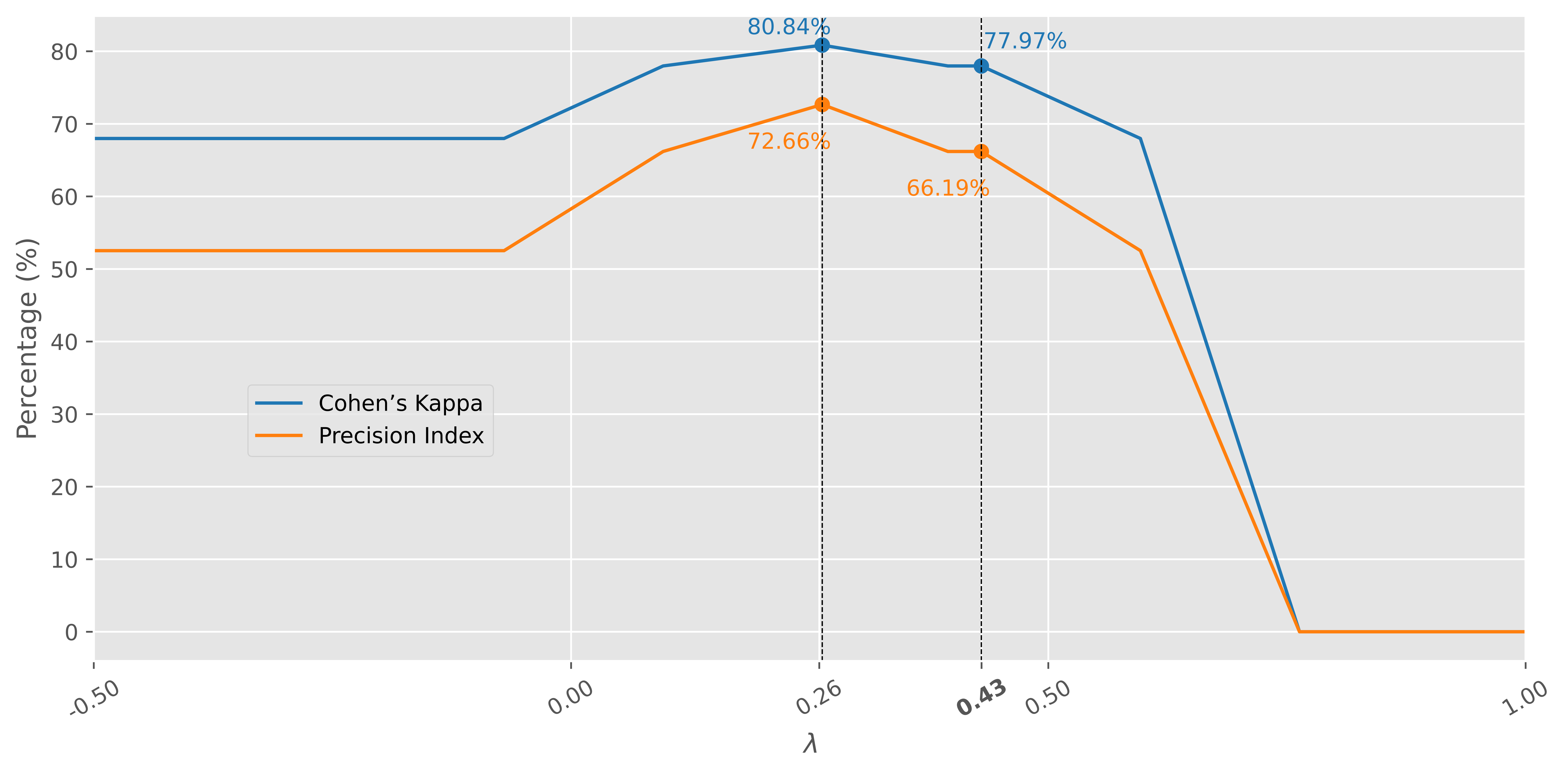}
    \caption{$\kappa$ and precision for the crack image as a function of $\lambda$. The value $\lambda=0.43$  maximizes the log-likelihood.}
    \label{fig:Grieta_Lambda}
\end{figure}

\begin{figure}[h!]
    \centering
    \includegraphics[scale=0.55]{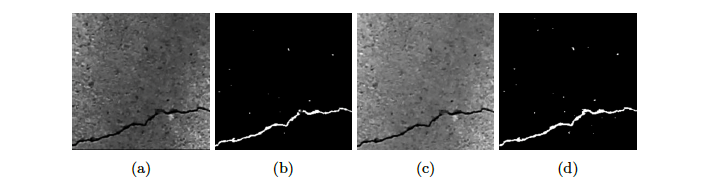}
    \caption{Transformed and predicted images. (a) Transformed image with $\lambda=0.43$; (b) Predicted image with $\lambda=0.43$; (c) Transformed image with $\lambda=0.26$; (d) Predicted image with $\lambda=0.26$.}.
    \label{fig:Grieta_Lambda2}
\end{figure}

We calculated the precision index and $\kappa$ between the mask and the prediction using the LDA method. Then we plotted the precision index and $\kappa$ versus $\lambda$ to explore the quality of segmentation as a function of the transformation parameter. The results of the crack image are shown in Figures \ref{fig:Grieta_Lambda}. The optimal value of $\lambda$, determined through likelihood, is observed to be neither a maximum for the concordance nor for the precision of the crack image.
Figure \ref{fig:Grieta_Lambda2} shows the transformed images and the predictions obtained using the $\lambda$ values that maximize concordance and precision of the crack image ($\lambda=0.31$), as well as the value that maximizes the log-likelihood ($\lambda=0.43$). In Figure \ref{fig:Grieta_Lambda2}, it can be seen that clarity improves significantly near the crack, highlighting details that were previously less visible and subsequently helping in image prediction.

\begin{figure}[h!]
    \centering
    \includegraphics[scale=0.5]{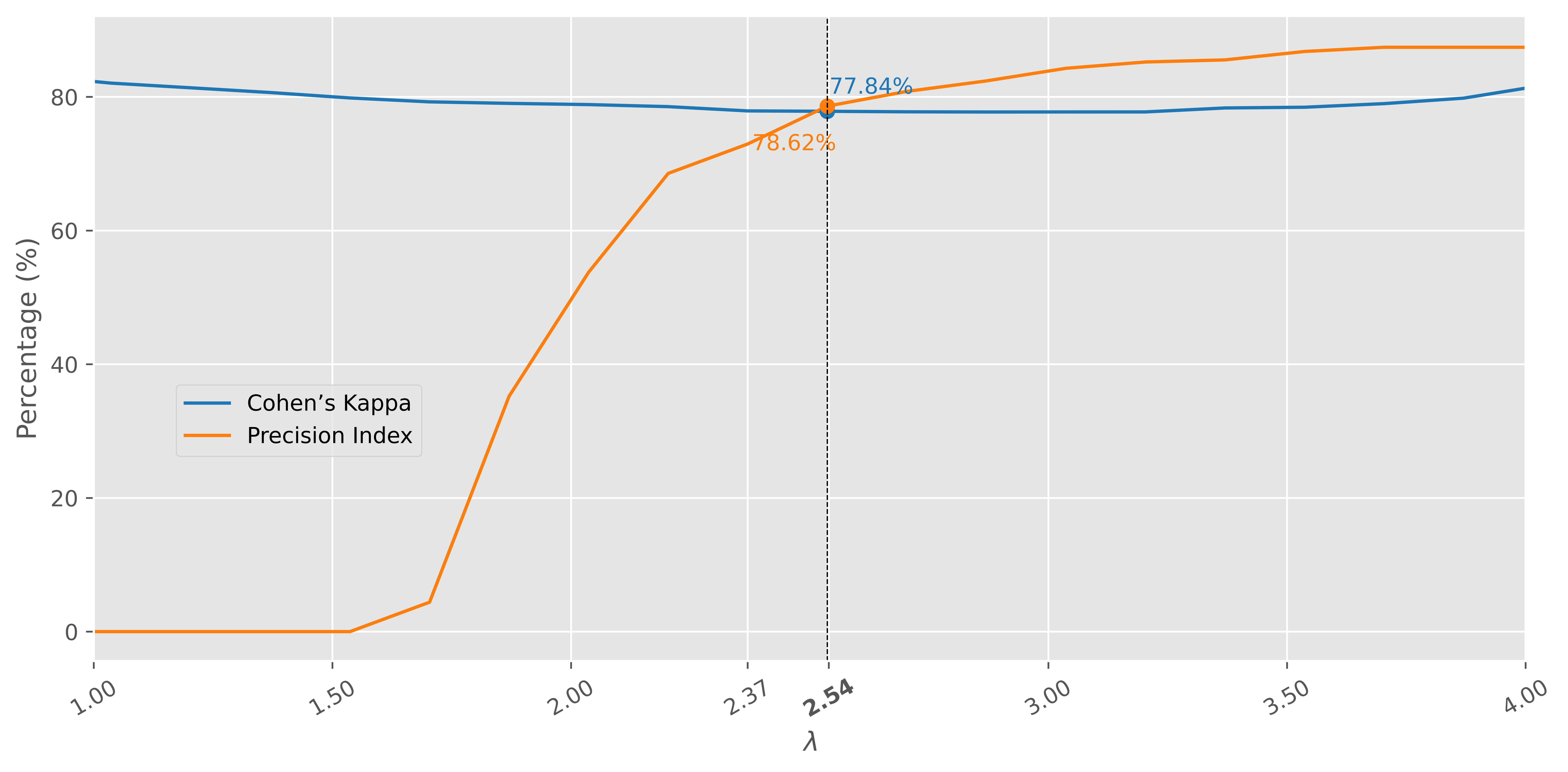}
    \caption{$\kappa$ and precision for the rock image as a function of $\lambda$. The value $\lambda = 2.54$ corresponds to the estimated parameter obtained from the Box-Cox transformation.}
    \label{fig:Rocas_Lambda}
\end{figure}

Similar to the crack image, Figure \ref{fig:Rocas_Lambda} illustrates the accuracy percentage for the lunar rock image and the concordance between the mask and the prediction using LDA. In this case, the point at which precision and $\kappa$ reach their maximum value coincides. Another notable finding is that the maximum and minimum of each curve correspond to the minimum and maximum of the other.

\begin{figure}[h!]
    \centering
    \includegraphics[scale=0.5]{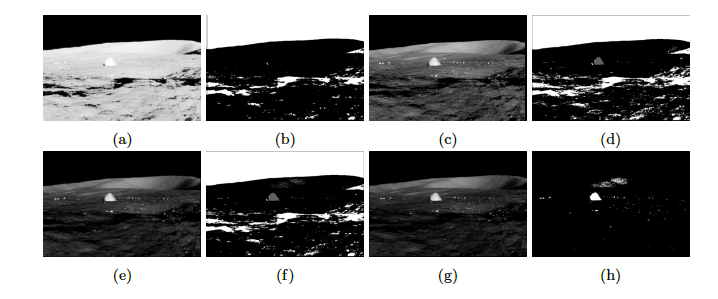}
    \caption{Transformed and predicted images. (a) Transformed image with $\lambda=0.43$; (b) Predicted image with $\lambda=0.43$; (c) Transformed image with $\lambda=2.54$; (d) Predicted image with $\lambda=2.54$; (e) Transformed image with $\lambda=4.04$; (f) Predicted image with $\lambda=4.04$; (g) Transformed image with $\lambda=4.54$; (h) Predicted image with $\lambda=4.54$.}
    \label{fig:Rocas_Lambda2}
\end{figure}

Figure \ref{fig:Rocas_Lambda2} displays images transformed using the Box-Cox transformation for various $\lambda$ values. As 
$\lambda$ varies, changes in contrast and image details become apparent, suggesting that a small 
$\lambda$ value, such as $\lambda=0.43$,
enhances overall visibility. Conversely, larger 
$\lambda$, like $\lambda=4.54$, emphasize bright areas, aiding in the identification of specific details. However, this increase in precision may come at the cost of lower agreement. Thus, selecting 
$\lambda$ involves a trade-off between precision and concordance.

%---------------------------
% Discussion
%---------------------------
\section{Discussion and final remarks}
This paper explored the application of the Box-Cox transformation as a preprocessing step for image segmentation, focusing on both deep learning and traditional machine learning methods. The results revealed varied impacts of the transformation across different models and image types.

For deep learning models, particularly U-Net, DeepLabV3, and FCN, the Box-Cox transformation did not lead to consistent improvements in segmentation performance. In fact, the application of the transformation often resulted in a decline in key metrics such as Dice and IoU. This outcome suggests that neural networks, which inherently learn complex representations from large amounts of data, do not significantly benefit from the transformation. However, in specific cases, such as the visual assessment of segmented images, the transformation enhanced image contrast and feature visibility.

In contrast, machine learning models showed more notable improvements after applying the transformation. Specifically, when segmenting crack images, LDA and QDA achieved up to a 3\% improvement across all performance metrics, demonstrating the transformation's effectiveness in enhancing feature separability. Confusion matrices confirmed that prefiltering helped these models better differentiate between object classes. In lunar rock image segmentation, QDA notably improved the classification of rock pixels, highlighting the transformation’s utility for models relying on distributional assumptions.

Another key finding of this study is the computational efficiency of traditional models. While deep learning approaches required extensive training, LDA and QDA achieved comparable or improved segmentation results with significantly lower computational costs. This efficiency is especially valuable in resource-constrained environments where computational power is limited.

The estimation of the transformation parameter $\lambda$ was also explored, demonstrating its importance in optimizing the transformation’s effect on image contrast and segmentation accuracy.  Further research in this area could incorporate the spatial autocorrelation of the image. This leads to consider a likelihood function with variance $\bm \Sigma(\bm \theta),$ where $\bm \theta$ is a parameter vector consisting of the variance components of a spatial model \cite{Cressie:1993}.
Another  avenue of research is exploring adaptive selection of  $\lambda$ and its effects on hybrid models that combine traditional and deep learning-based segmentation methods.

A compelling direction for continued research is the exploration of alternative techniques for symmetrizing distributions and achieving normality, such as the approach proposed by \cite{Yeo:2000}. Additionally, the effectiveness of Box-Cox-type transformations in combination with the wide range of prefiltering techniques available in the literature remains largely unexplored. In particular, a nonparametric version of the Box-Cox transformation has been applied to various regression models \cite{Zhou:2024}. This approach is particularly valuable for evaluating the distributional assumptions in spatial autoregressive models, especially when fitting a local window image for restoration purposes \cite{Ojeda:2010}. While the nonparametric Box-Cox method is model-based, it offers flexibility for application in both spatial and image processing contexts. Another promising direction for future research involves identifying appropriate methods for evaluating agreement between images. Recent advances in concordance measures for spatial data offer valuable tools for improving image comparisons by accounting for spatial correlation both within and between images \cite{Vallejos:2020}, \cite{Acosta:2024}. Additionally, extending evaluation metrics tailored to imbalanced data—following the approach of \cite{delaCruz:2025}—could provide more accurate assessments in contexts similar to the crack image analyzed in this study.

We recommend that practitioners and researchers first define their primary objective when applying prefiltering methods. For the Box-Cox transformation, selecting an appropriate $\lambda$ parameter is critical. It is advisable to explore a range of $\lambda$ values within a predefined grid to visually assess segmentation quality and ensure consistency with numerical metrics.

\backmatter

\bmhead{Supplementary information}

\bigskip
We provide a Supplementary Material with the  segmentation of an image yielded by QDA.

%---------------------------
% Acknoeledgements
%---------------------------
\bmhead{Acknowledgments}
This work was supported  by Fondecyt Grant  1230012,  
and by the AC3E, UTFSM, under grant AFB240002. 
R. Vallejos also acknowledges financial support from CONICYT through the STIC-AMSUD program, grant 23STIC-02.

%---------------------------
% Declarations
%---------------------------
\section*{Declarations}

{\bf Conflict of interest} 
The authors declare that they have no conflicts of interest related to this work.

\bigskip
\noindent {\bf Ethical approval}
All datasets and images used in this study are publicly available online.

\bigskip
\noindent {\bf Author contribution statements}
R. Vallejos and F. Osorio conceived and designed the study and methodology. Both contributed to the writing and revision of the manuscript. S. Vidal implemented the computational framework and prepared the initial draft. G. Britos contributed to the interpretation of results and manuscript revision. All authors reviewed and approved the final version of the paper.
%---------------------------
% Appendix A
%---------------------------
\begin{appendices}

\section{Segmentation Algorithms}\label{secA1}
\subsection{Support Vector Machine}

\bigskip

As discussed in Section \ref{subsec:PFI}, each image is represented as a vector \( \boldsymbol{x} \in \mathbb{R}_n^{+} \), where \( n \) is the number of pixels and the components of \( \boldsymbol{x} \) correspond to nonnegative intensity values.
In a pixel-based classification setting, each pixel is treated as a separate observation. 

Let \( \boldsymbol{x}_j \in \mathbb{R}^d \) denote a feature vector extracted from the \( j \)-th pixel, and let \( t_j \in \{-1, 1\} \) be its corresponding class label. We aim to classify a new pixel using a linear model of the form
\begin{equation}
    y(\boldsymbol{x}) = \boldsymbol{w}^T \phi(\boldsymbol{x}) + b,
    \label{eq:svm_pixel}
\end{equation}
where $ \phi(\boldsymbol{x}) $ is a transformation applied to the pixel feature vector (e.g., normalization or kernel embedding), and $b$ is a bias term. The decision is made according to the sign of $ y(\boldsymbol{x}) $.

Suppose the transformed feature vectors $ \phi(\boldsymbol{x}_j) $ are linearly separable. Then there exist parameters $ \boldsymbol{w} $ and $ b $ such that
$
    t_j y(\boldsymbol{x}_j) = t_j(\boldsymbol{w}^T \phi(\boldsymbol{x}_j) + b) > 0, \text{for all training pixels}.
$

To improve generalization, Support Vector Machines (SVM) find the separating hyperplane that maximizes the margin—the shortest distance from any pixel to the decision boundary. This distance is given by
$$
    \frac{t_j y(\boldsymbol{x}_j)}{\|\boldsymbol{w}\|} = \frac{t_j (\boldsymbol{w}^T \phi(\boldsymbol{x}_j) + b)}{\|\boldsymbol{w}\|}.
$$
The optimal parameters are obtained by solving
$$
    \arg\max_{\boldsymbol{w}, b} \left\{ \frac{1}{\|\boldsymbol{w}\|} \min_{j} \left[ t_j \left( \boldsymbol{w}^T \phi(\boldsymbol{x}_j) + b \right) \right] \right\}.
$$

In cases where pixel features are not linearly separable, the \emph{kernel trick} allows mapping them into a higher-dimensional space where linear separation is feasible. Rather than computing $ \phi(\boldsymbol{x}_i)^T \phi(\boldsymbol{x}_j) $ explicitly, a kernel function $K(\boldsymbol{x}_i, \boldsymbol{x}_j)$ is used to evaluate the similarity between two pixels directly in the original space. Common kernel functions for pixel-level classification are listed in Table~\ref{table:kernel_functions_pixel}.

\begin{table}[h]
\centering
\begin{tabular}{>{\bfseries}m{3.5cm} m{5cm} m{3cm}}
\toprule
\textbf{Kernel Name} & \textbf{Kernel Function} & \textbf{Parameters} \\
\midrule
Linear & $ K(\boldsymbol{x}, \boldsymbol{x}') = \boldsymbol{x}^T \boldsymbol{x}' $ & None \\
Polynomial & $K(\boldsymbol{x}, \boldsymbol{x}') = (\gamma \boldsymbol{x}^T \boldsymbol{x}' + r)^d $ & $ \gamma > 0 $, $ d \in \mathbb{Z}^+ $, $ r\geq 0 $ \\
Gaussian (RBF) & $ K(\boldsymbol{x}, \boldsymbol{x}') = \exp\left(-\gamma \|\boldsymbol{x} - \boldsymbol{x}'\|^2\right) $ & $ \gamma > 0 $ \\
Sigmoid & $ K(\boldsymbol{x}, \boldsymbol{x}') = \tanh(\gamma \boldsymbol{x}^T \boldsymbol{x}' + r)$ & $ \gamma > 0 $, $ r \in \mathbb{R} $ \\
\bottomrule
\end{tabular}
\caption{Common kernel functions used for pixel-based image classification with SVM}
\label{table:kernel_functions_pixel}
\end{table}

\subsection{Discriminant Analysis}

We consider a multi-class scenario with $K$ classes 
labeled by $ \{1, 2, \ldots, K\} $. Let $\pi_k$ denote the prior probability of class $k$,
with $\sum_{k=1}^{K} \pi_k = 1$. Furthermore, suppose $f_k(\boldsymbol{x})$ is the class-conditional 
density of the feature vector $\boldsymbol{x}$ given that the true class is $k$. By Bayes' theorem,
the posterior probability of class $k$ given $\boldsymbol{x}$ is
\begin{equation}
    P(G = k \,\big\vert\, \boldsymbol{x}) 
    = \dfrac{f_k(\boldsymbol{x})\,\pi_k}{\sum_{\ell = 1}^{K} f_{\ell}(\boldsymbol{x})\,\pi_{\ell}}.
    \label{eq:posterior_discriminant}
\end{equation}

A classical assumption in Discriminant Analysis is that each class $k$ can be modeled 
by a multivariate normal distribution:
$$
    f_k(\boldsymbol{x}) 
    = \dfrac{1}{(2\pi)^{d/2}\,\bigl|\boldsymbol{\Sigma}_k\bigr|^{1/2}}
      \exp\!\Bigl(-\tfrac{1}{2}\bigl(\boldsymbol{x}-\boldsymbol{\mu}_k\bigr)^\top 
                       \boldsymbol{\Sigma}_k^{-1}
                       \bigl(\boldsymbol{x}-\boldsymbol{\mu}_k\bigr)\Bigr),
$$
where $\boldsymbol{\mu}_k \in \mathbb{R}^d$ is the mean vector of class $k$ and
$\boldsymbol{\Sigma}_k \in \mathbb{R}^{d \times d}$ is its covariance matrix.

Linear Discriminant Analysis (LDA) arises under the additional assumption that 
all classes share the same covariance matrix $\boldsymbol{\Sigma}$, i.e., 
$\boldsymbol{\Sigma}_k = \boldsymbol{\Sigma}$ for all $k$. 
When comparing two classes \(k\) and \(\ell\), the classification decision 
depends on the log-ratio of their posterior probabilities:
\begin{align}
    \log \dfrac{P(G = k \mid \boldsymbol{x})}{P(G = \ell \mid \boldsymbol{x})}
    &= \log \dfrac{f_k(\boldsymbol{x})}{f_{\ell}(\boldsymbol{x})} 
       \;+\; \log \dfrac{\pi_k}{\pi_{\ell}} \nonumber \\[6pt]
    &= \log \dfrac{\pi_k}{\pi_{\ell}}
       \;-\; \tfrac{1}{2}\bigl(\boldsymbol{\mu}_k + \boldsymbol{\mu}_\ell\bigr)^\top 
                    \boldsymbol{\Sigma}^{-1}
                    \bigl(\boldsymbol{\mu}_k - \boldsymbol{\mu}_\ell\bigr)
       \;+\; \boldsymbol{x}^\top \boldsymbol{\Sigma}^{-1}
                     \bigl(\boldsymbol{\mu}_k - \boldsymbol{\mu}_\ell\bigr),
    \label{eq:lda_log_ratio}
\end{align}
which is linear in \(\boldsymbol{x}\). Hence, the decision boundary (where the two 
posteriors are equal) forms a hyperplane in $\mathbb{R}^d$. Extending this 
to all pairs of classes leads to purely linear boundaries that partition the feature 
space among the $K$ classes.

It is often convenient to define the \emph{linear discriminant functions}
$$
    \delta_k(\boldsymbol{x})
    \;=\;
    \boldsymbol{x}^\top \boldsymbol{\Sigma}^{-1}\boldsymbol{\mu}_k 
    \;-\;
    \tfrac{1}{2}\,\boldsymbol{\mu}_k^\top \boldsymbol{\Sigma}^{-1}\boldsymbol{\mu}_k 
    \;+\; 
    \log \pi_k,
$$
so the classification rule is
$$
    G(\boldsymbol{x}) 
    \;=\; 
    \arg\max_{1 \le k \le K} \, \delta_k(\boldsymbol{x}).
$$

In practice, the parameters \(\{\pi_k, \boldsymbol{\mu}_k, \boldsymbol{\Sigma}\}\) 
are unknown and estimated from the training set 
\(\{\boldsymbol{x}_i, g_i\}_{i=1}^N\), where \(g_i \in \{1,\dots,K\}\) is the class label of \(\boldsymbol{x}_i\). 
Typical estimates are
\begin{itemize}
    \item \(\widehat{\pi}_k = \tfrac{N_k}{N}\), \quad where \(N_k\) is the number of training pixels in class \(k\);
    \item \(\widehat{\boldsymbol{\mu}}_k 
            = \dfrac{1}{N_k}\!\sum_{g_i = k}\boldsymbol{x}_i;\)
    \item \(\widehat{\boldsymbol{\Sigma}} 
            = \dfrac{1}{N-K} \sum_{k=1}^K 
              \;\sum_{g_i = k} 
                  \bigl(\boldsymbol{x}_i - \widehat{\boldsymbol{\mu}}_k\bigr)
                  \bigl(\boldsymbol{x}_i - \widehat{\boldsymbol{\mu}}_k\bigr)^\top.\)
\end{itemize}

If \(\boldsymbol{\Sigma}_k\) are not constrained to be identical, then the 
class-conditional densities retain quadratic terms in \(\boldsymbol{x}\). 
In that case, the \emph{quadratic discriminant function} for class \(k\) is
\[
    \delta_k(\boldsymbol{x})
    \;=\;
    -\,\tfrac{1}{2}\,\log \bigl|\boldsymbol{\Sigma}_k\bigr|
    \;-\;\tfrac{1}{2}\,\bigl(\boldsymbol{x} - \boldsymbol{\mu}_k\bigr)^\top 
                     \boldsymbol{\Sigma}_k^{-1}\!
                     \bigl(\boldsymbol{x} - \boldsymbol{\mu}_k\bigr)
    \;+\; 
    \log \pi_k.
\]
As a result, the boundaries between classes become quadratic surfaces in \(\mathbb{R}^d\). 
Formally, each pairwise boundary for classes \(k\) and \(\ell\) is determined by
\(\{\boldsymbol{x} \,\mid\, \delta_k(\boldsymbol{x}) = \delta_\ell(\boldsymbol{x})\}\), 
which yields a quadratic equation in \(\boldsymbol{x}\).

\subsection{\texorpdfstring{$K$}{K}-Nearest Neighbors (KNN)}

A straightforward example of a non-parametric classifier is the 
\emph{$K$-Nearest Neighbors} ($K$-NN) algorithm. In the same spirit as the 
previous methods, let each pixel in the training set be described by a 
feature vector \(\boldsymbol{x}_i \in \mathbb{R}^d\) and a corresponding class 
label \(y_i \in \{1,2,\dots,K\}\). Denote the entire training set by 
\(\mathcal{D} = \{(\boldsymbol{x}_i,y_i)\}_{i=1}^N\).

Given a new pixel with feature vector \(\boldsymbol{x}\), the $K$-NN algorithm 
finds the set \(N_K(\boldsymbol{x},\mathcal{D}) \subseteq \{1,\dots,N\}\) of 
indices of the \(K\) closest training points to \(\boldsymbol{x}\) according 
to some distance metric \(d(\cdot,\cdot)\). The posterior probability for 
class \(c\) can then be estimated as the empirical frequency of class \(c\) 
among the \(K\) neighbors:
\begin{equation}
    P\bigl(G = c \,\big\vert\, \boldsymbol{x}\bigr) 
    \;=\; 
    \frac{1}{K}
    \sum_{i \,\in\, N_K(\boldsymbol{x},\mathcal{D})} 
    \mathbb{I}\!\bigl(y_i = c\bigr),
    \label{eq:knn_posterior}
\end{equation}
where \(\mathbb{I}\!\bigl(\cdot\bigr)\) is the indicator function, equal to 1 
if its argument is true and 0 otherwise. The predicted class is thus 
\[
    \widehat{G}(\boldsymbol{x}) 
    \;=\; 
    \arg\max_{1 \le c \le K} \; P\bigl(G = c \mid \boldsymbol{x}\bigr).
\]

The hyperparameter \(K\) plays a crucial role in balancing bias and variance. 
Choosing a very small \(K\) (e.g., \(K=1\)) makes the classifier sensitive to 
noisy or atypical points in the training data, leading to high variance and 
potential overfitting. On the other hand, choosing a large \(K\) may cause the 
classifier to smooth out subtle but informative local patterns, increasing bias 
and potentially underfitting by favoring the majority class in a broader region 
of the feature space.

The notion of ``nearest'' neighbors is determined by the distance metric
\(d(\mathbf{x},\mathbf{y})\). In this work we adopt the \emph{Euclidean distance}
$
  d(\mathbf{x},\mathbf{y}) \;=\;
  \left(\sum_{j=1}^{d} (x_j - y_j)^2\right)^{1/2},
$
whose simplicity, rotational invariance, and ubiquity in image–feature spaces
make it a natural baseline for measuring similarity.

The simplicity of $K$-NN makes it an attractive baseline for pixel-wise 
classification in image segmentation tasks. It places minimal assumptions on 
the underlying data distribution, but this comes at a computational cost 
during the prediction phase, since finding the nearest neighbors for a new 
\(\boldsymbol{x}\) can be expensive for large datasets.

\subsection{LightGBM (Light Gradient Boosting Machine)}

LightGBM is a supervised learning algorithm that applies a gradient boosting framework to 
decision trees, aiming to optimize both efficiency and predictive performance 
\cite{ke2017lightgbm}. Let \(\{(\boldsymbol{x}_i, y_i)\}_{i=1}^n\) be a training dataset, 
where \(\boldsymbol{x}_i \in \mathbb{R}^d\) is the feature vector for the \(i\)-th sample 
and \(y_i\) is its corresponding label or target value. The algorithm iteratively refines 
an ensemble of weak learners (decision trees) to minimize a global loss function. 
A typical loss function is:

\begin{equation}
    L(y_i, \hat{y}_i) \;=\; \sum_{i=1}^{n} \ell\bigl(y_i, \hat{y}_i\bigr),
    \label{eq:lightgbm_loss}
\end{equation}
\noindent
where $\ell(\cdot,\cdot)$ is the pointwise loss (e.g., quadratic loss or log-loss), 
$y_i$ is the true target, and $\hat{y}_i$ is the model prediction. 
LightGBM follows a \emph{gradient boosting} approach: at each iteration, it uses the 
negative gradients of the loss with respect to the current predictions 
($\hat{y}_i$) to create a new decision tree. Specifically, for each training point 
$(\boldsymbol{x}_i, y_i)$, the pseudo-residual (negative gradient) is defined as:

\begin{equation}
    g_i 
    \;=\; 
    \frac{\partial\,L\bigl(y_i, \hat{y}_i\bigr)}{\partial \,\hat{y}_i}.
    \label{eq:lightgbm_gradient}
\end{equation}

The next decision tree is then trained to approximate these gradient values 
$\{g_i\}$. To speed up training and reduce memory usage, LightGBM employs a 
leaf-wise tree growth strategy along with techniques such as histogram-based 
binning for feature values and gradient histograms for split-finding.

Unlike traditional \emph{level-wise} tree learners, which split all leaves in 
one level before going to the next, LightGBM splits the leaf that yields 
the largest reduction in the objective function first. This 
\emph{leaf-wise} approach often leads to deeper trees and faster loss reduction, 
although it may require hyperparameter tuning (e.g., $\max\_depth$) to avoid 
overfitting on certain datasets.

LightGBM partitions the data based on maximizing an information gain criterion. 
In broad terms, a split on subset $S$ of the training data is accepted if it 
increases the following quantity:

\begin{equation}
    \mathrm{Gain}(S) 
    \;=\; 
    \sum_{j} \Bigl(\frac{\lvert S_j\rvert}{\lvert S\rvert} 
                   \cdot \mathrm{Inf}(S_j)\Bigr),
    \label{eq:lightgbm_gain}
\end{equation}

\noindent
where $S$ is the current node’s dataset, $\{S_j\}$ are the child nodes resulting 
from the split, and $\mathrm{Inf}(\cdot)$ is a measure of impurity (e.g., 
\emph{entropy} or \emph{Gini index}). In regression problems, the splitting criterion 
is typically based on variance reduction or a similar numeric measure.

\subsection{U-Net: Convolutional Networks}
U-Net \cite{ronneberger2015unet} was introduced in 2015. Originally designed for biomedical image segmentation, this neural network architecture has also proven effective in a wide range of other fields, including satellite image segmentation, agriculture, and construction. Its structure consists of two main phases: encoding and decoding, forming a symmetric network shaped like a `U'.

In the encoding phase, the input image is progressively downsampled through convolutional filters to extract relevant features. In the decoding phase, the feature representation follows an inverse path, gradually increasing its dimensions back to the original image size, where the final output is the generated segmentation mask. A key difference from the traditional \textit{encoder--decoder} structure is the presence of skip connections that transfer feature maps from the encoding path to the decoding path. These connections facilitate mask generation and improve gradient propagation, preventing vanishing gradients as the network is traversed backward.

For a more detailed description of the U-Net architecture and its functioning, the reader is encouraged to consult the original paper \cite{ronneberger2015unet}.

\subsection{DeepLabv3+}
DeepLabV3+ is an extension of the DeepLab family of architectures aimed at 
improving semantic image segmentation \cite{Chen:2018}. 
Its design follows an \textit{encoder--decoder} approach, where the \textit{encoder} 
is typically based on a modified Xception network. This encoder leverages 
\textit{depthwise separable} and \textit{atrous} convolutions to enlarge the receptive 
field without increasing the number of parameters. A central component of the encoder 
is the Atrous Spatial Pyramid Pooling (ASPP) module, which applies parallel atrous 
convolutions with different dilation rates. This strategy allows the model to capture 
multi-scale contextual information and effectively segment objects of varying sizes.

The \textit{decoder} stage fuses high-level features from the ASPP module with 
lower-level (yet spatially more precise) features extracted at earlier layers 
of the network. A subsequent \textit{upsampling} operation restores the spatial 
resolution, enhancing the delineation of object boundaries and preserving fine details. 
Finally, each pixel is assigned a semantic label, producing a high-resolution 
segmentation map where object contours are more accurately defined than in 
previous versions of DeepLab.

For a more detailed understanding of the model's architecture and functioning, 
the reader is encouraged to consult the original paper \cite{Chen:2018}.

\section{Lunar Rock Images and Tables}\label{secA2}

\begin{figure}[h]
    \centering
    \begin{subfigure}[b]{0.32\linewidth}
    \includegraphics[width=\linewidth]{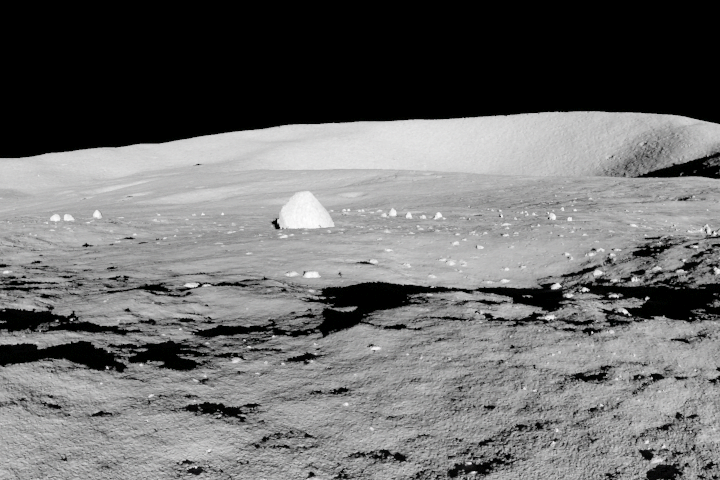}
    \caption{}
    \end{subfigure}
    \begin{subfigure}[b]{0.32\linewidth}
    \includegraphics[width=\linewidth]{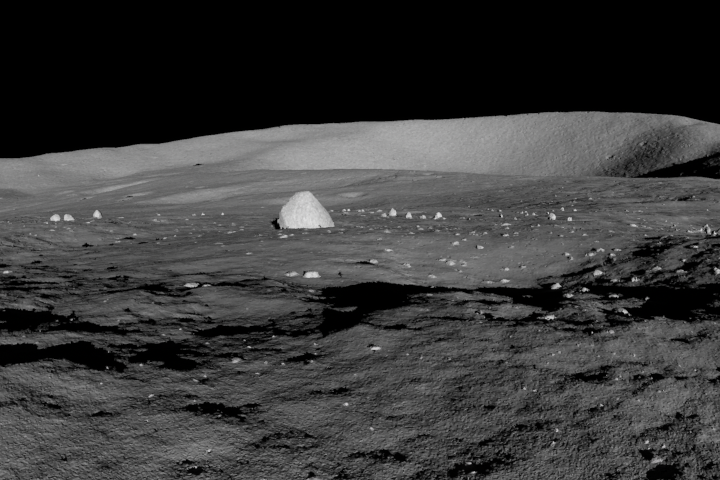}
    \caption{}
    \end{subfigure}
    \begin{subfigure}[b]{0.32\linewidth}
    \includegraphics[width=\linewidth]{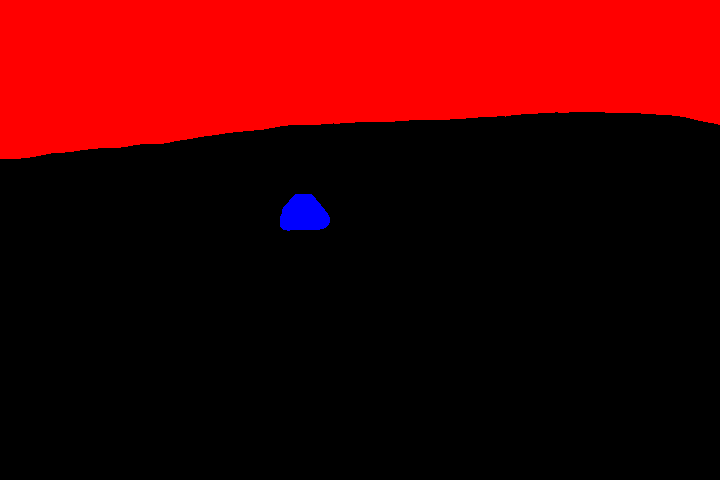}
    \caption{}
    \end{subfigure}
    \caption{Rock lunar images. (a) Original image; (b) Image transformed using the Box-Cox method; (c) Mask image.}
    \label{fig:luna_modelos}
\end{figure}

\begin{table}[htp]
\centering
\scalebox{1.3}{
\begin{tabular}{lccc}
\toprule
Method               & Precision     & Recall        & $F_1$ score   \\ \midrule
SVM                  & 96          & 96          & 96         \\
SVM Box-Cox & 95 & 96 & 95 \\ 
LDA      & 93         & 92         & 92          \\ 
LDA Box-Cox & 93 & 92 & 92 \\ 
QDA                  & 95          & 95          & 95          \\ 
QDA Box-Cox & 94 & 93 &93 \\ \bottomrule
\end{tabular}
}
\caption{Quality assessment (percentage of correct segmentation) of machine learning models.}
\label{table:metodos_luna}
\end{table}

\begin{table}[h]
\centering
\scalebox{1.3}{
\begin{tabular}{lcccccc}
\toprule
 & \multicolumn{3}{c}{Before Transformation} & \multicolumn{3}{c}{After Transformation}\\
 \cmidrule(lr){2-4} \cmidrule(lr){5-7}
                  & Sky & Rock & Lunar Surface & Sky & Rock & Lunar Surface\\ \midrule
Sky & 91             & 0   &9   & 89   & 0   & 11      \\ 
Rock  & 100             & 0 & 0 & 21             & 79 & 0  \\ 
Lunar surface   & 1             & 0  & 99 & 1    & 0  & 99 \\ 
 \bottomrule
\end{tabular}
}
\caption{Percentages of normalized confusion matrices for the segmentation results obtained using QDA before and after applying the Box-Cox transformation for the lunar rock.}
\label{table:antes_despues_roca}
\end{table}

\begin{table}[h]
\centering
\scalebox{1.3}{
\begin{tabular}{ccccc}
\toprule
Method & SVM & LightGBM & LDA  & QDA\\ \midrule
Time (s) & 5298 &  4458&  6 & 4.02 \\ \bottomrule
\end{tabular}
}
\caption{Training times for the estimation of hyperparameters for each of the methods used in the segmentation of the lunar rock image.}
\label{table:times_rock}
\end{table}

\end{appendices}

\end{document}